\documentclass[preprint,12pt]{elsarticle}

\usepackage{amsmath}
\usepackage{amssymb}
\usepackage{hyperref}
\usepackage[hyphenbreaks]{breakurl}
\usepackage[capitalize]{cleveref}
\usepackage{booktabs}
\usepackage{url}
\usepackage{subcaption}
\usepackage[utf8]{inputenc}
\usepackage[T1]{fontenc}

\biboptions{sort&compress}

\newcounter{bla}

\journal{Computer Physics Communications}

\makeatletter
\def\printFirstPageNotes{%
  \ifx\@cornotes\@empty\else\@cornotes\fi
   \begin{flushright}
     {\tt DESY 18-064},
     {\tt KA-TP-09-2018}
   \end{flushright}
}
\makeatother

\begin{document}

\begin{frontmatter}

\title{\texttt{N2HDECAY}: Higgs Boson Decays in the Different Phases of the N2HDM}

\author[a]{Isabell Engeln}
\author[a]{Margarete Mühlleitner\corref{maggie}}
\author[b]{Jonas Wittbrodt\corref{jonas}}

\cortext[maggie] {milada.muhlleitner@kit.edu}
\cortext[jonas] {jonas.wittbrodt@desy.de}

\address[a]{Institute for Theoretical Physics, Karlsruhe Institute of Technology, Wolfgang-Gaede-Str. 1, 76131 Karlsruhe, Germany}
\address[b]{Theory Group, Deutsches Elektronen-Synchrotron DESY, Notkestr. 85, 22607 Hamburg, Germany}

\begin{abstract}
The program \texttt{N2HDECAY} calculates the branching ratios and decay widths of the Higgs bosons of the Next-to-Two-Higgs-Doublet  Model (N2HDM).
The code incorporates the dominant higher-order effects by including QCD corrections and off-shell decay modes.
The N2HDM is an extension of the Standard Model by a Higgs doublet and a real Higgs singlet.
Its phenomenology can change dramatically depending on which global symmetries are broken by electroweak symmetry breaking.
It can feature a large visible Higgs sector in the broken phase, behave like a 2HDM with scalar singlet dark matter in the dark singlet phase, or, in the inert doublet phase, extend an inert doublet model by mixing a singlet with the SM Higgs boson.
\texttt{N2HDECAY} provides precise predictions for the decays of the Higgs bosons in all of these phases.
\end{abstract}

\end{frontmatter}

% Computer program descriptions should contain the following
% PROGRAM SUMMARY.
\newpage
{\bf PROGRAM SUMMARY}\\
\begin{small}
\noindent
{\em Program Title:} \texttt{N2HDECAY} \\
{\em Licensing provisions:} MIT  \\
{\em Programming language:} Fortran  \\

{\em A code to calculate the branching ratios and total widths of the Higgs bosons in the N2HDM including state-of-the-art QCD and off-shell corrections. The code is available at }\url{https://gitlab.com/jonaswittbrodt/N2HDECAY}\,.
\end{small}

\section{Introduction}
Experimental studies conducted since the discovery of the 125\,GeV Higgs boson~\cite{Aad:2012tfa,Chatrchyan:2012xdj} have established it to behave Standard Model (SM)-like.
The resulting constraints on models beyond the SM (BSM) force minimal extensions of the Higgs sector into specific corners of the parameter space where new-physics effects in the 125\,GeV Higgs boson are small.
This makes minimal extensions of the SM scalar sector, such as the Two-Higgs-Doublet Model (2HDM)~\cite{Lee:1973iz,Branco:2011iw}, less useful for experimental analysis since potentially interesting collider signatures are likely to be incompatible with other observations.
Therefore, non-minimal extensions of the Higgs sector become more appealing since they can account for interesting novel collider signatures while remaining compatible with existing measurements.

The Next-to-Two-Higgs-Doublet Model (N2HDM)~\cite{Chen:2013jvg,Drozd:2014yla,Muhlleitner:2016mzt} adds a real, gauge-singlet scalar field to the Higgs sector of the 2HDM.
If this singlet acquires a vacuum expectation value (VEV) it mixes with the two CP-even Higgs bosons of the 2HDM.
The resulting enlargement of the parameter space with interesting collider signatures has been studied in~\cite{Muhlleitner:2016mzt} using the first published version of the program \texttt{N2HDECAY}.
The N2HDM is similar to the 2HDM plus singlet (2HDM+S) framework~\cite{Curtin:2013fra} where a complex singlet scalar is added to an aligned 2HDM. However, in the N2HDM we do not assume the 125\,GeV Higgs boson to be exactly aligned.

An additional feature of non-minimal extensions of the Higgs sector is that they may both modify the visible phenomenology of the model and add dark matter (DM) candidates.
Both dark singlet DM (for a recent study see~\cite{Athron:2017kgt}) and inert doublet DM~\cite{Deshpande:1977rw,Ma:2006km,LopezHonorez:2006gr,Belyaev:2016lok,Ilnicka:2015jba} can be realized within the N2HDM as different phases of electroweak symmetry breaking (EWSB).
In the dark singlet phase (DSP) the singlet field does not acquire a VEV and becomes a DM candidate.
In the inert doublet phase (IDP) one of the doublets has no VEV and forms a dark sector.
A phenomenological study of the dark phases of the N2HDM using \texttt{N2HDECAY} has been performed recently ~\cite{Engeln:2018ywp}.

The code \texttt{N2HDECAY} calculates branching ratios and total widths of the Higgs bosons in these three phases of the N2HDM.
It includes state-of-the-art higher-order QCD corrections and off-shell effects taken over from its parent code \texttt{HDECAY}~\cite{Djouadi:1997yw,Butterworth:2010ym,Djouadi:2018xqq}.
\texttt{N2HDECAY} therefore provides precise predictions for the phenomenology of the N2HDM Higgs bosons.
In this manual we first define the N2HDM with its three phases in \cref{sec:model}. Afterwards in \cref{sec:corrections}, we discuss the higher-order corrections included in the code. In \cref{sec:usage} we give user instructions for the code, describe the input and output formats as well as additional tools distributed with the code.
Finally, \cref{app:output} lists all decay modes calculated in \texttt{N2HDECAY} and their position in the output files while \cref{app:DSP} and \cref{app:IDP} contain the expressions for the parameter transformations and triple-Higgs couplings in the DSP and IDP, respectively.

\section{The Next-to Two Higgs Doublet Model}\label{sec:model}
The N2HDM \cite{Chen:2013jvg,Drozd:2014yla,Muhlleitner:2016mzt} extends the SM by a second Higgs doublet and a real Higgs singlet. The resulting scalar potential for the doublets $\Phi_{1,2}$ and the singlet $\Phi_S$ reads
\begin{equation}
  \begin{aligned}
  V &= m_{11}^2 |\Phi_1|^2 + m_{22}^2 |\Phi_2|^2 - m_{12}^2 (\Phi_1^\dagger
  \Phi_2 + \text{h.c.}) + \frac{\lambda_1}{2} (\Phi_1^\dagger \Phi_1)^2 +
  \frac{\lambda_2}{2} (\Phi_2^\dagger \Phi_2)^2  \\
  &\hphantom{=} + \lambda_3
  (\Phi_1^\dagger \Phi_1) (\Phi_2^\dagger \Phi_2) + \lambda_4
  (\Phi_1^\dagger \Phi_2) (\Phi_2^\dagger \Phi_1) + \frac{\lambda_5}{2}
  [(\Phi_1^\dagger \Phi_2)^2 + \text{h.c.}]  \\
  &\hphantom{=} + \frac{m_S^2}{2}  \Phi_S^2 + \frac{\lambda_6}{8} \Phi_S^4 +
  \frac{\lambda_7}{2} (\Phi_1^\dagger \Phi_1) \Phi_S^2 +
  \frac{\lambda_8}{2} (\Phi_2^\dagger \Phi_2) \Phi_S^2 \,,
\end{aligned}
  \label{eq:n2hdmpot}
\end{equation}
where all quartic couplings $\lambda_{1,\ldots,8}$ and mass parameters $m_{11,22,12,S}^2$ are real.
The first two lines of \cref{eq:n2hdmpot}, containing only the doublet fields, correspond to the standard 2HDM~\cite{Branco:2011iw}. The last line contains the contributions of the singlet field.

We impose two $\mathbb{Z}_2$ parities on the scalar potential \cref{eq:n2hdmpot}.
The symmetry called $\mathbb{Z}_2$,
\begin{align}
  \Phi_1\to\Phi_1\,,\ \Phi_2\to-\Phi_2\,,\ \Phi_S\to\Phi_S\,, \label{eq:Z2}
\end{align}
is the trivial generalization of the usual 2HDM $\mathbb{Z}_2$ symmetry to the N2HDM.
It can be extended to the Yukawa sector to prevent flavor changing neutral currents.
It is softly broken by $m_{12}^2\neq0$ and spontaneously broken if $\Phi_2$ develops a non-zero VEV.
The second symmetry, $\mathbb{Z}'_2$,
\begin{align}
  \Phi_1\to\Phi_1\,,\ \Phi_2\to\Phi_2\,,\ \Phi_S\to-\Phi_S\,, \label{eq:Z2p}
\end{align}
is spontaneously broken if the VEV of $\Phi_S$ is non-zero.
If any of these two parities remain unbroken after EWSB they lead to a dark sector containing a stable dark matter candidate.

After EWSB we expand the Higgs fields around their VEVs $v_1$, $v_2$ and $v_s$ as
\begin{equation}
\Phi_{i=\lbrace 1,2\rbrace} = \begin{pmatrix} \phi_i^+ \\ \frac{1}{\sqrt{2}} (v_i +
    \rho_i + i \eta_i) \end{pmatrix}\,,\quad
\Phi_S = v_S + \rho_S \,. \label{eq:n2hdmfields}
\end{equation}
The minimum conditions for the electroweak vacuum lead to the relations
\begin{align}
v_2 m_{12}^2 - v_1 m_{11}^2 &= \frac{1}{2} (v_1^3 \lambda_1 +
v_2^2 v_1 \lambda_{345} + v_S^2 v_1 \lambda_7)\,, \label{eq:n2hdmmin1} \\
v_1 m_{12}^2 - v_2 m_{22}^2 &= \frac{1}{2} (v_1^2 v_2 \lambda_{345} +
v_2^3 \lambda_2 + v_S^2 v_2 \lambda_8)\,, \label{eq:n2hdmmin2} \\
- v_S m_S^2 &= \frac{1}{2} v_S (v_1^2 \lambda_7 + v_2^2 \lambda_8 + v_S^2
\lambda_6)\,,  \label{eq:n2hdmmin3}
\end{align}
with
\begin{equation}
\lambda_{345} \equiv \lambda_3 + \lambda_4 + \lambda_5 \,.
\label{eq:l345}
\end{equation}
We have not canceled any powers of VEVs in \cref{eq:n2hdmmin1,eq:n2hdmmin2,eq:n2hdmmin3} to allow for the possibility of vanishing $v_{1,2,S}$.
These conditions can be used to express the mass parameters $m_{11}^2$, $m_{22}^2$ and $m_{S}^2$ through the VEVs and allow for four different phases depending on which VEVs are non-zero.

\subsection{The Broken Phase}
In the broken phase all CP- and charge-conserving VEVs ($v_1$, $v_2$ and $v_S$) are non-zero. This breaks both $\mathbb{Z}_2$ and $\mathbb{Z}'_2$. Like in the 2HDM, the rotation into the mass basis with the matrix
\begin{align}
  R_\beta = \begin{pmatrix}
  \cos\beta &  \sin\beta\\
  -\sin\beta & \cos\beta
\end{pmatrix}
\end{align}
yields one physical CP-odd Higgs boson $A$ and the neutral Goldstone boson $G^0$ from $\eta_{1,2}$ as well as a pair of charged Higgs bosons $H^\pm$ and charged Goldstone bosons $G^\pm$ from $\phi^\pm_{1,2}$. However, $v_S\neq0$ leads to a mixing between the $\rho_{1,2}$ and $\rho_S$ resulting in three physical CP-even Higgs bosons
\begin{equation}
\begin{pmatrix} H_1 \\ H_2 \\ H_3 \end{pmatrix} = R \begin{pmatrix}\rho_1 \\ \rho_2 \\ \rho_S\end{pmatrix}\,.
\end{equation}
The $3\times3$ orthogonal mixing matrix $R$ is parametrized by three angles $\alpha_{1,2,3}$ as
\begin{equation}
  R = \begin{pmatrix}
  c_{\alpha_1} c_{\alpha_2} & s_{\alpha_1} c_{\alpha_2} & s_{\alpha_2}\\
  -(c_{\alpha_1} s_{\alpha_2} s_{\alpha_3} + s_{\alpha_1} c_{\alpha_3})
  & c_{\alpha_1} c_{\alpha_3} - s_{\alpha_1} s_{\alpha_2} s_{\alpha_3}
  & c_{\alpha_2} s_{\alpha_3} \\
  - c_{\alpha_1} s_{\alpha_2} c_{\alpha_3} + s_{\alpha_1} s_{\alpha_3} &
  -(c_{\alpha_1} s_{\alpha_3} + s_{\alpha_1} s_{\alpha_2} c_{\alpha_3})
  & c_{\alpha_2}  c_{\alpha_3}
  \end{pmatrix}\,,
  \label{eq:mixingmatrix}
\end{equation}
where we introduced the short-hand notations $s_x\equiv\sin x$ and $c_x\equiv\cos x$.
We assume that the three mass eigenstates $H_{1,2,3}$ are ordered as $m_{H_1}<m_{H_2}<m_{H_3}$.

\begin{table}[bt]
  \caption{Coupling coefficients $c(H_i ff)$ of the Yukawa couplings of
     the N2HDM Higgs bosons $H_i$. \label{tab:yukcoup}}
\centering
  \begin{tabular}{rccc} \toprule
& $u$-type & $d$-type & leptons \\ \midrule
type I & $\frac{R_{i2}}{s_\beta}$
& $\frac{R_{i2}}{s_\beta}$ &
$\frac{R_{i2}}{s_\beta}$ \\
type II & $\frac{R_{i2}}{s_\beta} $
& $\frac{R_{i1}}{c_\beta} $ &
$\frac{R_{i1}}{c_\beta} $ \\
lepton-specific & $\frac{R_{i2}}{s_\beta}$
& $\frac{R_{i2}}{s_\beta}$ &
$\frac{R_{i1}}{c_\beta}$ \\
flipped & $\frac{R_{i2}}{s_\beta}$
& $\frac{R_{i1}}{c_\beta}$ &
$\frac{R_{i2}}{s_\beta}$ \\ \bottomrule
\end{tabular}
\end{table}

The three CP-even Higgs bosons couple to the SM gauge bosons $V\in\{W^\pm,Z\}$ with an effective coupling normalized to the SM given by
\begin{equation}
  c(H_i VV)=R_{i 1}\cos\beta + R_{i2}\sin\beta\,.\label{eq:gaugecoupbroken}
\end{equation}
To prevent tree-level flavor-changing neutral currents we extend the symmetry $\mathbb{Z}_2$, \cref{eq:Z2}, to the Yukawa sector. As in the 2HDM~\cite{Branco:2011iw}, this results in the well-known four distinct Yukawa types.
The effective couplings for the different types normalized to their SM values are given in \cref{tab:yukcoup}.
The couplings of $A$ and $H^\pm$ to SM particles are the same as in the 2HDM and can be found in the corresponding literature~\cite{Branco:2011iw}.
All triple-Higgs couplings needed for the computation of the decay widths are given in the appendix of \cite{Muhlleitner:2016mzt}.

In the broken phase we use the input parameters
\begin{equation}
  \alpha_{1,2,3}\,, \quad \tan\beta \,, \quad v \, ,
  \quad v_S \, , \quad m_{H_{1,2,3}} \,, \quad m_A \,, \quad m_{H^\pm}
  \,, \quad m_{12}^2 \,. \label{eq:brokeninputpars}
\end{equation}
The parameters $\tan\beta$ and $v$ are given by
\begin{align}
  \tan\beta=\frac{v_2}{v_1}\,,\quad v=v_1^2+v_2^2\approx 246\, \text{GeV}\,,\label{eq:tbetadef}
\end{align}
where the value of $v$ is the SM VEV.
The relations between this parameter set and the parameters of the scalar potential~\cref{eq:n2hdmpot} can be found in the appendix of~\cite{Muhlleitner:2016mzt}.

\subsection{The Dark Singlet Phase}
The dark singlet phase (DSP) is characterized by $v_S=0$ such that $\mathbb{Z}'_2$, \cref{eq:Z2p}, remains exact after EWSB. Any mixing between $\rho_{1,2}$ and $\rho_S$ is therefore forbidden and the CP-even mass eigenstates are
\begin{equation}
  \begin{pmatrix} H_1 \\ H_2 \\ H_D\\ \end{pmatrix} = R \begin{pmatrix}\rho_1 \\ \rho_2 \\ \rho_S\\\end{pmatrix}
\end{equation}
with
\begin{equation}
  R=\begin{pmatrix}
    -\sin\alpha &\cos\alpha &0\\
    \cos\alpha &\sin\alpha &0\\
    0&0&1\\
  \end{pmatrix}\,.\label{eq:mixmatdsp}
\end{equation}
The mixing matrix $R$ is chosen such that the mixing angle $\alpha$ follows the conventions for the 2HDM~\cite{Branco:2011iw} if we identify $h\equiv H_1$ and $H\equiv -H_2$.
For consistency with the other phases we will call the Higgs bosons $H_1$ and $H_2$ throughout this paper and assume them to be ordered as $m_{H_1}<m_{H_2}$.
The singlet scalar $H_D$ is a DM candidate stabilized by $\mathbb{Z}'_2$.
The charged and CP-odd sector are identical to the broken phase. The DSP of the N2HDM is equivalent to adding scalar singlet DM to a 2HDM.

The couplings of the Higgs bosons to SM particles in the DSP can be obtained from \cref{eq:gaugecoupbroken} and \cref{tab:yukcoup} by using  the definition \cref{eq:mixmatdsp}. The triple-Higgs couplings differ from the broken phase since all vertices involving an odd number of $H_D$ are forbidden by $\mathbb{Z}'_2$. The expressions for the relevant couplings can be found in \cref{app:tripleHDSP}.

In the DSP we cannot express all of the quartic couplings of the scalar potential \cref{eq:n2hdmpot} through masses and mixing angles. The two quartic Higgs-portal couplings $\lambda_7$ and $\lambda_8$ as well as the singlet self-coupling $\lambda_6$ remain free parameters. However, $\lambda_6$ does not enter any of the three-particle vertices relevant for Higgs decays and can be ignored for the purposes of this code. We therefore use the input parameters
\begin{equation}
  \alpha\,, \quad \tan\beta \,, \quad v \, ,
  \quad m_{H_{1,2}}\,, \quad m_{H_D}\,, \quad m_A \,, \quad m_{H^\pm}
  \,, \quad m_{12}^2\,,\quad \lambda_{7,8}  \,, \label{eq:dspinputpars}
\end{equation}
with $\tan\beta$ and $v$ defined through \cref{eq:tbetadef}.
The transformations from these parameters to the parameters of \cref{eq:n2hdmpot} are given in \cref{app:transDSP}.

\subsection{The Inert Doublet Phase}
The inert doublet phase (IDP) can only be realized if $m_{12}^2=0$ such that $\mathbb{Z}_2$, \cref{eq:Z2}, is not explicitly broken. If additionally $v_2=0$ the doublet $\Phi_2$ is inert~\cite{Deshpande:1977rw,Ma:2006km,LopezHonorez:2006gr,Belyaev:2016lok,Ilnicka:2015jba}.
The component fields of $\Phi_2$ are all mass eigenstates and form a dark sector consisting of a CP-even scalar $H_D$, a CP-odd scalar $A_D$ and a charged scalar $H^\pm_D$.
The only mixing in the IDP is between the singlet field $\rho_S$ and the doublet field $\rho_1$. This results in two visible CP-even neutral Higgs bosons $H_{1,2}$ through
\begin{equation}
  \begin{pmatrix} H_1 \\ H_2 \\ H_D\\ \end{pmatrix} = R \begin{pmatrix}\rho_1 \\ \rho_2 \\ \rho_S\\\end{pmatrix}
\end{equation}
with
\begin{equation}
  R=\begin{pmatrix}
    \cos\alpha &  0 &\sin\alpha\\
    -\sin\alpha & 0 &\cos\alpha\\
    0&1&0\\
  \end{pmatrix}\,,\label{eq:mixmatIDP}
\end{equation}
where we again take $m_{H_1}<m_{H_2}$.

The IDP is equivalent to extending an inert doublet model by a real singlet which mixes with the SM Higgs boson. The couplings of the $H_{i=1,2}$ to any pair of SM particles $x$ are therefore simply rescaled by
\begin{equation}
  c(H_i xx) = R_{i1}\,.
\end{equation}
Since this applies to the Yukawa couplings as well there are no different Yukawa types in the IDP.
The triple-Higgs couplings of the IDP can be found in \cref{app:tripleHIDP}.

 In the IDP \cref{eq:n2hdmmin2} is trivially satisfied so that we cannot express $m_{22}^2$ through the other parameters.
 Additionally, the inert doublet self-coupling $\lambda_2$ and the portal coupling $\lambda_8$ cannot be expressed through masses and mixing angles.
 However, $\lambda_2$ does not appear in any of the vertices relevant for Higgs boson decays and is not needed by the code. The set of required input parameters for the IDP are therefore
\begin{equation}
  \alpha \,,\quad v \,,\quad v_S \,,\quad m_{H_{1,2}} \,,\quad m_{H_D} \,,\quad m_{A_D} \,,\quad m_{H^\pm_D} \,,\quad m_{22}^2 \,,\quad \lambda_8\,,\label{eq:idpinputpars}
\end{equation}
where $v_1=v$ is the SM VEV.
The relations between the parameters of \cref{eq:idpinputpars} and \cref{eq:n2hdmpot} are given in \cref{app:transIDP}.

\subsection{The SM+DM Phase}
If $m_{12}^2=0$ and both $v_2$ and $v_S$ vanish the N2HDM becomes the SM extended by both an inert doublet and a scalar DM singlet. This phase is currently not implemented in the code.

\section{Higgs Decays in N2HDECAY}\label{sec:corrections}
The code \texttt{N2HDECAY} is an extension of the 2HDM implementation in \texttt{HDECAY}~\cite{Djouadi:1997yw,Butterworth:2010ym,Harlander:2013qxa,Djouadi:2018xqq} to the N2HDM.
The calculation of the decay widths includes the most relevant QCD corrections since they factorize and can be taken over from the SM or the MSSM.
The SM electroweak corrections, on the other hand, cannot be easily adapted to models with extended scalar sectors and are not taken into account. They have been recently provided in~\cite{Krause:2017mal}.
The currently implemented higher-order corrections are
\begin{itemize}
  \item massive NLO corrections near threshold~\cite{Braaten:1980yq,Sakai:1980fa,Inami:1980qp,Drees:1989du,Drees:1990dq} and massless $\mathcal{O}(\alpha_s^4)$ corrections far above threshold~\cite{Gorishnii:1983cu,Gorishnii:1990zu,Gorishnii:1991zr,Kataev:1993be,Surguladze:1994gc,Larin:1995sq,Melnikov:1995yp,Chetyrkin:1995pd,Chetyrkin:1996sr,Baikov:2005rw} to decays into quark pairs.
  \item $\text{N}^3\text{LO}$ corrections in the heavy-top-limit to decays into $gg$~\cite{Inami:1982xt,Djouadi:1991tka,Spira:1993bb,Spira:1995rr,Chetyrkin:1997iv,Chetyrkin:1997un,Kramer:1996iq,Chetyrkin:1998mw,Schroder:2005hy,Chetyrkin:2005ia,Baikov:2006ch}.
  \item QCD corrections to the quark loops in the loop-mediated ($W$, heavy-fermion and charged-Higgs-boson loops) decay into $\gamma\gamma$ including the full quark-mass dependence~\cite{Zheng:1990qa,Djouadi:1990aj,Djouadi:1993ji,Melnikov:1993tj,Dawson:1992cy,Inoue:1994jq,Spira:1995rr,Fleischer:2004vb,Harlander:2005rq}.
\end{itemize}
The QCD corrections to quark loops in the loop-induced $Z\gamma$ decay mode are numerically small~\cite{Spira:1991tj,Bonciani:2015eua,Gehrmann:2015dua} and are not taken into account.

The code additionally includes decays into off-shell gauge bosons and top-quarks where appropriate. This includes
\begin{itemize}
  \item decays into a pair of off-shell massive gauge bosons~\cite{Cahn:1988ru},
  \item off-shell gauge bosons in all decays into a Higgs boson and a gauge boson~\cite{Moretti:1994ds,Djouadi:1995gv},
  \item off-shell $t$-quarks in decays into $t$-quark pairs and the $t\bar{b}$, $t\bar{s}$ and $t\bar{d}$ decay modes of the charged Higgs boson $H^+$~\cite{Moretti:1994ds,Djouadi:1995gv}.
\end{itemize}
\texttt{N2HDECAY} does not include off-shell Higgs bosons in decays into pairs of Higgs bosons.
These off-shell corrections are only available in \texttt{HDECAY} assuming that the off-shell Higgs boson decays predominantly into $b\bar{b}$.
This is justified for the 125\,GeV Higgs boson but may be a bad approximation for other Higgs bosons of the N2HDM.
Off-shell corrections are therefore disabled for all decays into Higgs boson pairs.
Finally, the code also calculates the branching ratios and total width of the top-quark to account for possibly allowed decays into $H^+b$.
A complete list of all decay channels calculated by \texttt{N2HDECAY} including off-shell decays is given in \cref{tab:BP_output,tab:DSP_output,tab:IDP_output}.

\section{User Instructions}\label{sec:usage}
The \texttt{N2HDECAY} source code is available at
\begin{center}
  \url{https://gitlab.com/jonaswittbrodt/N2HDECAY}\,.
\end{center}
The compilation of the code requires only a Fortran compiler. The code can be compiled either with the supplied \texttt{Makefile}, in which the correct compiler has to be selected manually, or using \texttt{cmake}.
The code compiles into a single executable called \texttt{n2hdecay}.
It reads input from a file called \texttt{n2hdecay.in} in the current working directory and writes its output files containing the branching ratios and total widths into the current directory. Input file and output folder may be changed by providing arguments to the optional \texttt{-i} and \texttt{-o} command line flags, respectively.

The input file is a plain text file containing the parameters of the model. The first parameter to be specified is the \verb/PHASE/ which is 1 for the broken phase, 2 for the DSP and 3 for the IDP. The parameter \verb/YUKTYPE/ specifies the Yukawa type. The possible values are 1 (Type I), 2 (Type II), 3 (lepton-specific) and 4 (flipped).
In the next block the parameters $\tan\beta$, $m_{12}^2$, $m_A$, $m_{H^\pm}$ and $v_S$ can be specified. This block is used for every phase. If the value for one of the parameters is fixed by the phase the value given here is ignored.
After these initial blocks of common parameters there is a dedicated block containing the remaining parameters for each phase.
The final block of the input file contains the SM parameters defined as in \texttt{HDECAY} (see~\cite{Djouadi:2018xqq}).
An overview of the required N2HDM parameters for each phase is shown in \cref{tab:inputpars}.
The SM VEV $v$, which is an input parameter in every phase (see \cref{eq:brokeninputpars,eq:dspinputpars,eq:idpinputpars}), is calculated internally from the value of $G_F$ given in the SM parameter block.

\begin{table}[tb]
  \caption{Required input parameters in the different phases. See also \cref{eq:brokeninputpars,eq:dspinputpars,eq:idpinputpars}. Parameters with a -- are fixed by the phase and the value in the input file is ignored.}
  \label{tab:inputpars}
  \begin{subtable}[t]{0.32\textwidth}
  \caption{broken phase}
  \centering
  \begin{tabular}{l c}
    \toprule
    Parameter & Value \\
    \midrule
    \verb/PHASE/ & 1 \\
    \verb/YUKTYPE/ & 1,2,3,4 \\
    \midrule
    \verb/tbeta/ & $\tan\beta$ \\
    \verb/m12sq/ & $m_{12}^2$ \\
    \verb/m_A/ & $m_A$ \\
    \verb/m_Hc/ & $m_{H^\pm}$ \\
    \verb/vs/ & $v_S$\\
    \midrule
    \multicolumn{2}{c}{PHASE = 1 block}\\
    \midrule
    \verb/m_H1/ & $m_{H_1}$\\
    \verb/m_H2/ & $m_{H_2}$\\
    \verb/m_H3/ & $m_{H_3}$\\
    \verb/a1/ & $\alpha_1$\\
    \verb/a2/ & $\alpha_2$\\
    \verb/a3/ & $\alpha_3$\\
    \bottomrule
  \end{tabular}
\end{subtable}
\begin{subtable}[t]{0.32\textwidth}
  \caption{DSP}
  \centering
  \begin{tabular}{l c}
    \toprule
    Parameter & Value \\
    \midrule
    \verb/PHASE/ & 2 \\
    \verb/YUKTYPE/ & 1,2,3,4 \\
    \midrule
    \verb/tbeta/ & $\tan\beta$ \\
    \verb/m12sq/ & $m_{12}^2$ \\
    \verb/m_A/ & $m_A$ \\
    \verb/m_Hc/ & $m_{H^\pm}$ \\
    \verb/vs/ & -- \\
    \midrule
    \multicolumn{2}{c}{PHASE = 2 block}\\
    \midrule
    \verb/m_H1/ & $m_{H_1}$\\
    \verb/m_H2/ & $m_{H_2}$\\
    \verb/m_HD/ & $m_{H_D}$\\
    \verb/alpha/ & $\alpha$\\
    \verb/L7/ & $\lambda_7$\\
    \verb/L8/ & $\lambda_8$\\
    \bottomrule
  \end{tabular}
\end{subtable}
\begin{subtable}[t]{0.32\textwidth}
  \caption{IDP}
  \centering
  \begin{tabular}{l c}
    \toprule
    Parameter & Value \\
    \midrule
    \verb/PHASE/ & 3 \\
    \verb/YUKTYPE/ & -- \\
    \midrule
    \verb/tbeta/ & -- \\
    \verb/m12sq/ & -- \\
    \verb/m_A/ & $m_{A_D}$ \\
    \verb/m_Hc/ & $m_{H^\pm_D}$ \\
    \verb/vs/ & $v_S$\\
    \midrule
    \multicolumn{2}{c}{PHASE = 3 block}\\
    \midrule
    \verb/m_H1/ & $m_{H_1}$\\
    \verb/m_H2/ & $m_{H_2}$\\
    \verb/m_HD/ & $m_{H_D}$\\
    \verb/alpha/ & $\alpha$\\
    \verb/L8/ & $\lambda_8$\\
    \verb/m22sq/ & $m_{22}^2$\\
    \bottomrule
  \end{tabular}
\end{subtable}
\end{table}

Running the code produces a set of output files containing the branching ratios and decay widths of all Higgs bosons and the top-quark.
The files are named \verb!br.Hx_N2HDM_y! where \verb!Hx! denotes the decaying Higgs bosons.
For neutral \verb!Hx!=\texttt{H1,H2,H3,A} the files with \texttt{y}=\texttt{a} contain the decays into SM fermions, \texttt{y}=\texttt{b} the decays into SM gauge bosons and \texttt{y}=\texttt{c,d} the decays into double-Higgs and gauge-Higgs final states as well as the total width.
The decays of the charged Higgs boson \texttt{Hx=H+} into fermions are given in the \texttt{y}=\texttt{a,b} files and \texttt{y}=\texttt{c} contains the decays into gauge-Higgs final states and the total width.
The gauge-Higgs decay modes of the dark-sector particles in the IDP \texttt{Hx=HD,AD,HD+} along with their total widths can be found in the files \verb!br.Hx_N2HDM!.
The mass of the decaying particle is included in every file as the first entry.
The file \verb!br.top! contains the mass of the charged Higgs boson, the top-quark branching ratios and its total width.
Each of the files includes a header specifying the content.
For a complete overview of the output files and their contents in the different phases see \cref{tab:BP_output,tab:DSP_output,tab:IDP_output}.

For the user's convenience we also provide a Python 3 script
\begin{center}
  \verb@tools/IO_N2HDECAY.py@
\end{center}
with the utility functions:
\begin{itemize}
  \item \verb/IO_N2HDECAY.write(filepath, phase, params)/: Generates an input file for \texttt{N2HDECAY} at \texttt{filepath} for the given \verb/phase/ and the parameters taken from the Python dictionary \verb/params/.
  \item \verb/IO_N2HDECAY.read(brfolder)/: Reads the results of an \texttt{N2HDECAY}  run from the \texttt{brfolder} and returns them as a Python dictionary.
\end{itemize}

In the subfolder \texttt{tests} we provide three example input files (one for each phase) with the corresponding results of the code. The folder also contains a Python 3 script \texttt{testN2HDECAY.py} which can be used to compare the program output against these example results. Invoking \texttt{make test} when compiling with \texttt{cmake} automatically calls this script for each phase.

Finally, the source code also includes an up-to-date version of this manual and a \texttt{CHANGELOG} file detailing all changes in the code since the initial release.

\newpage
\appendix

\section{Decay Modes and Output Format}\label{app:output}
The \cref{tab:BP_output,tab:DSP_output,tab:IDP_output} in this section list all decay modes calculated by \texttt{N2HDECAY} and the content of the output files in all three phases.
\begin{table}[!h]
  \caption{The branching ratios calculated in the broken phase and their location in the output files. Particles marked with $^{(*)}$ are allowed to be off-shell.}\label{tab:BP_output}
  \centering
  \setlength{\tabcolsep}{1.5pt}
  \begin{tabular}{lcccccc}
    \toprule
    file & \multicolumn{6}{c}{$H_1 \to$}  \\
    \midrule
    \verb/br.H1_N2HDM_a/ & $b\bar{b}$ & $\tau\bar{\tau}$ &$\mu\bar{\mu}$ & $s\bar{s}$ & $c\bar{c}$ & $t^{(*)}\bar{t}^{(*)}$ \\
    \verb/br.H1_N2HDM_b/ & $g g$ & $\gamma\gamma$ & $Z\gamma$ & ${W^\pm}^{(*)}{W^\mp}^{(*)}$ & $Z^{(*)}Z^{(*)}$\\
    \verb/br.H1_N2HDM_c/ & $A A$ & $Z^{(*)} A$ & ${W^\pm}^{(*)} H^\mp$ & $H^\pm H^\mp$ & $\Gamma_\text{tot}$\\
    \midrule
    & \multicolumn{6}{c}{$H_2\to$}\\
    \midrule
    \verb/br.H2_N2HDM_a/ & $b\bar{b}$ & $\tau\bar{\tau}$ &$\mu\bar{\mu}$ & $s\bar{s}$ & $c\bar{c}$ & $t^{(*)}\bar{t}^{(*)}$ \\
    \verb/br.H2_N2HDM_b/ & $g g$ & $\gamma\gamma$ & $Z\gamma$ & ${W^\pm}^{(*)}{W^\mp}^{(*)}$ & $Z^{(*)}Z^{(*)}$\\
    \verb/br.H2_N2HDM_c/ & $H_1 H_1$ & $A A$ & $Z^{(*)} A$ & ${W^\pm}^{(*)} H^\mp$ & $H^\pm H^\mp$ & $\Gamma_\text{tot}$\\
    \midrule
    & \multicolumn{6}{c}{$H_3\to$}\\
    \midrule
    \verb/br.H3_N2HDM_a/ & $b\bar{b}$ & $\tau\bar{\tau}$ &$\mu\bar{\mu}$ & $s\bar{s}$ & $c\bar{c}$ & $t^{(*)}\bar{t}^{(*)}$ \\
    \verb/br.H3_N2HDM_b/ & $g g$ & $\gamma\gamma$ & $Z\gamma$ & ${W^\pm}^{(*)}{W^\mp}^{(*)}$ & $Z^{(*)}Z^{(*)}$\\
    \verb/br.H3_N2HDM_c/ & $H_1 H_1$ & $H_1 H_2$ & $H_2 H_2$ & $A A$ & $Z^{(*)} A$ & ${W^\pm}^{(*)} H^\mp$\\
    \verb/br.H3_N2HDM_d/  & $H^\pm H^\mp$ & $\Gamma_\text{tot}$ \\
    \midrule
    & \multicolumn{6}{c}{$A \to$}  \\
    \midrule
    \verb/br.A_N2HDM_a/ & $b\bar{b}$ & $\tau\bar{\tau}$ &$\mu\bar{\mu}$ & $s\bar{s}$ & $c\bar{c}$ & $t^{(*)}\bar{t}^{(*)}$ \\
    \verb/br.A_N2HDM_b/ & $g g$ & $\gamma\gamma$ & $Z\gamma$ \\
    \verb/br.A_N2HDM_c/ & $Z^{(*)}H_1$ & $Z^{(*)}H_2$ & $Z^{(*)}H_3$ & ${W^\pm}^{(*)} H^\mp$ & $\Gamma_\text{tot}$\\
    \midrule
    & \multicolumn{6}{c}{$H^{+} \to$}  \\
    \midrule
    \verb/br.H+_N2HDM_a/ & $c\bar{b}$ & $\tau\bar{\nu_\tau}$ &$\mu\bar{\nu_\mu}$ & $u\bar{s}$ & $c\bar{s}$ & $t^{(*)}\bar{b}$ \\
    \verb/br.H+_N2HDM_b/ & $c\bar{d}$ & $u\bar{b}$ & $t^{(*)}\bar{s}$ & $t^{(*)}\bar{d}$ \\
    \verb/br.H+_N2HDM_c/ & ${W^+}^{(*)} H_1$ & ${W^+}^{(*)} H_2$ & ${W^+}^{(*)} H_3$ & ${W^+}^{(*)} A$ & $\Gamma_\text{tot}$\\
    \midrule
    & \multicolumn{6}{c}{$t \to$}  \\
    \midrule
    \verb/br.top/ & $W^+ b$ & $H^+ b$ & $\Gamma_\text{tot}$\\
    \bottomrule
  \end{tabular}
\end{table}

\begin{table}
  \caption{The branching ratios calculated in the DSP and their location in the output files. Particles marked with $^{(*)}$ are allowed to be off-shell.}\label{tab:DSP_output}
  \centering
  \setlength{\tabcolsep}{1.5pt}
  \begin{tabular}{lcccccc}
    \toprule
    file & \multicolumn{6}{c}{$H_1 \to$}  \\
    \midrule
    \verb/br.H1_N2HDM_a/ & $b\bar{b}$ & $\tau\bar{\tau}$ &$\mu\bar{\mu}$ & $s\bar{s}$ & $c\bar{c}$ & $t^{(*)}\bar{t}^{(*)}$ \\
    \verb/br.H1_N2HDM_b/ & $g g$ & $\gamma\gamma$ & $Z\gamma$ & ${W^\pm}^{(*)}{W^\mp}^{(*)}$ & $Z^{(*)}Z^{(*)}$\\
    \verb/br.H1_N2HDM_c/ & $H_D H_D$ & $A A$ & $Z^{(*)} A$ & ${W^\pm}^{(*)} H^\mp$ & $H^\pm H^\mp$ & $\Gamma_\text{tot}$\\
    \midrule
    & \multicolumn{6}{c}{$H_2\to$}\\
    \midrule
    \verb/br.H2_N2HDM_a/ & $b\bar{b}$ & $\tau\bar{\tau}$ &$\mu\bar{\mu}$ & $s\bar{s}$ & $c\bar{c}$ & $t^{(*)}\bar{t}^{(*)}$ \\
    \verb/br.H2_N2HDM_b/ & $g g$ & $\gamma\gamma$ & $Z\gamma$ & ${W^\pm}^{(*)}{W^\mp}^{(*)}$ & $Z^{(*)}Z^{(*)}$\\
    \verb/br.H2_N2HDM_c/ & $H_D H_D$ & $H_1 H_1$ & $A A$ & $Z^{(*)} A$ & ${W^\pm}^{(*)} H^\mp$ & $H^\pm H^\mp$\\
    \verb/br.H2_N2HDM_d/ & $\Gamma_\text{tot}$\\
    \midrule
    & \multicolumn{6}{c}{$A \to$}  \\
    \midrule
    \verb/br.A_N2HDM_a/ & $b\bar{b}$ & $\tau\bar{\tau}$ &$\mu\bar{\mu}$ & $s\bar{s}$ & $c\bar{c}$ & $t^{(*)}\bar{t}^{(*)}$ \\
    \verb/br.A_N2HDM_b/ & $g g$ & $\gamma\gamma$ & $Z\gamma$  \\
    \verb/br.A_N2HDM_c/ & $Z^{(*)}H_1$ & $Z^{(*)}H_2$ & ${W^\pm}^{(*)} H^\mp$ & $\Gamma_\text{tot}$\\
    \midrule
    & \multicolumn{6}{c}{$H^{+} \to$}  \\
    \midrule
    \verb/br.H+_N2HDM_a/ & $c\bar{b}$ & $\tau\bar{\nu_\tau}$ &$\mu\bar{\nu_\mu}$ & $u\bar{s}$ & $c\bar{s}$ & $t^{(*)}\bar{b}$ \\
    \verb/br.H+_N2HDM_b/ & $c\bar{d}$ & $u\bar{b}$ & $t^{(*)}\bar{s}$ & $t^{(*)}\bar{d}$\\
    \verb/br.H+_N2HDM_c/ & ${W^+}^{(*)} H_1$ & ${W^+}^{(*)} H_2$ & ${W^+}^{(*)} A$ & $\Gamma_\text{tot}$\\
    \midrule
    & \multicolumn{6}{c}{$t \to$}  \\
    \midrule
    \verb/br.top/ & $W^+ b$ & $H^+ b$ & $\Gamma_\text{tot}$\\
    \bottomrule
  \end{tabular}
\end{table}

\begin{table}
  \caption{The branching ratios calculated in the IDP and their location in the output files. Particles marked with $^{(*)}$ are allowed to be off-shell.}\label{tab:IDP_output}
  \centering
  \setlength{\tabcolsep}{1.5pt}
  \begin{tabular}{lcccccc}
    \toprule
    file & \multicolumn{6}{c}{$H_1 \to$}  \\
    \midrule
    \verb/br.H1_N2HDM_a/ & $b\bar{b}$ & $\tau\bar{\tau}$ &$\mu\bar{\mu}$ & $s\bar{s}$ & $c\bar{c}$ & $t^{(*)}\bar{t}^{(*)}$ \\
    \verb/br.H1_N2HDM_b/ & $g g$ & $\gamma\gamma$ & $Z\gamma$ & ${W^\pm}^{(*)}{W^\mp}^{(*)}$ & $Z^{(*)}Z^{(*)}$\\
    \verb/br.H1_N2HDM_c/ & $H_D H_D$ & $A_D A_D$ & $Z^{(*)} A_D$ & ${W^\pm}^{(*)} H_D^\mp$ & $H_D^\pm H_D^\mp$ & $\Gamma_\text{tot}$\\
    \midrule
    & \multicolumn{6}{c}{$H_2\to$}\\
    \midrule
    \verb/br.H2_N2HDM_a/ & $b\bar{b}$ & $\tau\bar{\tau}$ &$\mu\bar{\mu}$ & $s\bar{s}$ & $c\bar{c}$ & $t^{(*)}\bar{t}^{(*)}$ \\
    \verb/br.H2_N2HDM_b/ & $g g$ & $\gamma\gamma$ & $Z\gamma$ & ${W^\pm}^{(*)}{W^\mp}^{(*)}$ & $Z^{(*)}Z^{(*)}$\\
    \verb/br.H2_N2HDM_c/ & $H_D H_D$ & $H_1 H_1$ & $A_D A_D$ & $Z^{(*)} A_D$ & ${W^\pm}^{(*)} H_D^\mp$ & $H_D^\pm H_D^\mp$ \\
    \verb/br.H2_N2HDM_d/ & $\Gamma_\text{tot}$\\
    \midrule
    & \multicolumn{6}{c}{$H_D \to$}  \\
    \midrule
    \verb/br.HD_N2HDM/ & $Z^{(*)}A_D$ & ${W^\pm}^{(*)} H_D^\mp$ & $\Gamma_\text{tot}$ \\
    \midrule
    & \multicolumn{6}{c}{$A_D \to$}  \\
    \midrule
    \verb/br.AD_N2HDM/ & $Z^{(*)}H_D$ & ${W^\pm}^{(*)} H_D^\mp$ & $\Gamma_\text{tot}$ \\
    \midrule
    & \multicolumn{6}{c}{$H_D^{+} \to$}  \\
    \midrule
    \verb/br.HD+_N2HDM/ & ${W^+}^{(*)} H_D$ & ${W^+}^{(*)} A_D$ & $\Gamma_\text{tot}$\\
    \midrule
    & \multicolumn{6}{c}{$t \to$}  \\
    \midrule
    \verb/br.top/ & $W^+ b$ & $H^+ b$ & $\Gamma_\text{tot}$\\
    \bottomrule
  \end{tabular}
\end{table}

\section{Parameter Transformations and Triple-Higgs Couplings in the DSP}\label{app:DSP}
In this section we present the transformations between the parameter set \cref{eq:dspinputpars} and the parameters of \cref{eq:n2hdmpot} in the DSP. We also provide the formulas for the triple-Higgs couplings.
These expressions have first been derived in~\cite{Engeln:2018ywp}.

\subsection{Parameter Transformations}\label{app:transDSP}
We first use the minimum conditions \cref{eq:n2hdmmin1,eq:n2hdmmin2} to express the mass parameter $m_{11}^2$ and $m_{22}^2$ as
\begin{align}
m_{11}^2 =&  \tan\beta m_{12}^2 -\frac{v^2}{2} \left(c_\beta^2 \lambda_1 +  s_\beta^2 \lambda_{345}\right)\,,\\
m_{22}^2  =& \cot\beta m_{12}^2-\frac{v^2}{2}\left(c_\beta^2 \lambda_{345} + s_\beta^2 \lambda_2 \right)\,,
\end{align}
where we use the short-hand notations \cref{eq:l345}, $s_x\equiv\sin x$ and $c_x\equiv\cos x$.
To obtain the remaining transformations we solve the linear system given by the rotations into the mass basis. This yields
\begin{align}
m_{S}^2 =&  -\frac{1}{2}\left(v_1^2 \lambda_7 + v_2^2 \lambda_8 - 2 m_{H_D}\right)\,,\\
\lambda_1 =& \frac{1}{v^2 c^2_\beta}\left[\left(\sum_{i=1}^2 m_{H_{i}}^2 R^2_{i1}\right) -m_{12}^2 \frac{s_\beta}{c_\beta}\right]\,,\\
\lambda_2 =& \frac{1}{v^2 s^2_\beta}\left[\left(\sum_{i=1}^2 m_{H_{i}}^2 R^2_{i2}\right) -m_{12}^2 \frac{c_\beta}{s_\beta}\right]\,,\\
\lambda_3 =& \frac{1}{v^2 c_\beta s_\beta}\left[\left(\sum_{i=1}^2 m_{H_{i}}^2 R_{i1} R_{i2}\right) - m_{12}^2\right] + \frac{2}{v^2} m_{H^\pm}^2\,,\\
\lambda_4 =& \frac{1}{v^2}\left(m_A^2 - 2 m_{H^\pm}^2\right) + \frac{1}{v^2 c_\beta s_\beta} m_{12}^2\,,\\
\lambda_5 =& - \frac{1}{v^2} m_A^2 + \frac{1}{v^2 c_\beta s_\beta}m_{12}^2\,,
\end{align}
with $R_{ij}$ given by \cref{eq:mixmatdsp}.

\subsection{Triple-Higgs Couplings}\label{app:tripleHDSP}
The triple-Higgs couplings are defined through
\begin{align}
  g(X_i X_j X_k)=\left.\frac{\partial^3\mathcal{L}}{\partial X_i \partial X_j \partial X_k}\right|_\text{VEV}\,,\label{eq:tripleHdef}
\end{align}
where in the DSP $X_{i,j,k} \in \left\{H_1, H_2, A, H^\pm, H_D\right\}$.
The couplings have first been derived in~\cite{Engeln:2018ywp}.
Using $H_{i,j}\in\{H_1,H_2\}$, $R_{ij}$ from \cref{eq:mixmatdsp} and $\lambda_{34-5}\equiv\lambda_3+\lambda_4-\lambda_5$ they are given by
\begin{align}
g(H_iH_iH_i) &=
\begin{aligned}[t]
3v \Big[
&c_\beta \left(
R^3_{i1} \lambda_1 + R_{i1}R^2_{i2} \lambda_{345} \right)\\
&+ s_\beta \left(
R^3_{i2} \lambda_2 + R_{i2}R^2_{i1} \lambda_{345}
\right)
\Big]\,,
\end{aligned}\\
g(H_iH_jH_j) &=
\begin{aligned}[t]
v \Big[
&c_\beta \left(
3 R_{i1} R^2_{j1} \lambda_1
+ (3 R_{i2}  R_{j1} R_{j2} + R_{i1}) \lambda_{345}
\right)\\
&+ v s_\beta \left(
3 R_{i2} R^2_{j2} \lambda_2
+ (3 R_{i1} R_{j1} R_{j2} + R_{i2}) \lambda_{345}
\right)\Big]\,,
\end{aligned}\\
g(H_iAA) &=
\begin{aligned}[t]
v \Big[&
c_\beta\left(
c_\beta s_\beta R_{i2} \left(\lambda_2 -2\lambda_5\right)
+ c^2_\beta R_{i1} \lambda_{34-5}
\right)\\
&+ s_\beta \left(
c_\beta s_\beta R_{i2} \left(\lambda_1 -2 \lambda_5\right)
+ s^2_\beta R_{i2} \lambda_{34-5}
\right)
\Big]\,,
\end{aligned}\\
g(H_iH^+H^-) &=
\begin{aligned}
v \Big[
& c_{\beta} \left(
s^2_\beta R_{i1} \lambda_1
+ c^2_\beta R_{i1} \lambda_3
- c_{\beta} s_{\beta} R_{i2} \left(\lambda_4 + \lambda_5\right)
\right)\\
&+ s_\beta \left(
 c^2_\beta R_{i2} \lambda_2
+ s^2_\beta R_{i2} \lambda_3
- c_{\beta} s_{\beta} R_{i1} \left(\lambda_4 + \lambda_5\right)
\right)
\Big]\,,
\end{aligned}\\
g(H_iH_DH_D) &= v\Big[c_\beta R_{i1} \lambda_7 + s_\beta R_{i2} \lambda_8\Big]\,.
\end{align}
All other trilinear couplings among the $X_i$ vanish.

\section{Parameter Transformations and Triple-Higgs Couplings in the IDP}\label{app:IDP}
The relations between the parameters of the Lagrangian \cref{eq:n2hdmpot} and the input parameter set \cref{eq:idpinputpars} for the IDP as well as the triple-Higgs couplings were derived in~\cite{Engeln:2018ywp}.
In~\cite{Engeln:2018ywp} the charges of $\Phi_1$ and $\Phi_2$ under the $\mathbb{Z}_2$ symmetry are switched. Their convention can be restored from ours by replacing
\begin{align}
  m_{11}^2\leftrightarrow m_{22}^2\,,\quad \lambda_1\leftrightarrow \lambda_2\,,\quad \lambda_7\leftrightarrow\lambda_8\,,\quad R_{i1}\leftrightarrow R_{i2}
\end{align}
in all formulas given below.

\subsection{Parameter Transformations}\label{app:transIDP}
Using the minimum conditions \cref{eq:n2hdmmin1,eq:n2hdmmin3} we obtain
\begin{align}
m_{11}^2 &= -\frac{1}{2}\left(v^2\lambda_1+v_S^2\lambda_7\right),\\
m_{S}^2 &= -\frac{1}{2}\left(v_S^2\lambda_6+v^2\lambda_7\right).
\end{align}
We can then solve for the parameters of the Lagrangian and obtain
\begin{align}
\lambda_1 =& \frac{1}{v^2}\left(\sum_{i=1}^2 m_{H_{i}}^2  R^2_{i1}\right)\,,\\
\lambda_3 =& \frac{1}{v^2}\left(2\left(m_{H^{\pm}_D}^2 - m_{11}^2\right) - v_s^2\, \lambda_8\right)\,, \\
\lambda_4 =& \frac{1}{v^2}\left(m_{A_D}^2 + m_{H_{D}}^2 - 2  m_{H^{\pm}}^2\right)\,,\\
\lambda_5 =& \frac{1}{v^2}\left(m_{H_{D}}^{2}-m_{A_D}^{2}\right)\,,\\
\lambda_6 =& \frac{1}{v_s^2}\left(\sum_{i=1}^2 m_{H_{i}}^2  R^2_{i3}\right)\,,\\
\lambda_7 =& \frac{1}{v v_s}\left(\sum_{i=1}^2 m_{H_{i}}^2  R_{i1}  R_{i3}\right)\,,
\end{align}
with $R_{ij}$ given by \cref{eq:mixmatIDP}.

\subsection{Triple-Higgs Couplings}\label{app:tripleHIDP}
The triple-Higgs couplings of the IDP are defined through \cref{eq:tripleHdef} with $X_{i,j,k} \in \left\{H_1, H_2, H_D, A_D, H^\pm_D\right\}$. Due to the $\mathbb{Z}_2$ parity any couplings with an odd number of inert particles vanish. The non-zero triple-Higgs couplings are given by
\begin{align}
  g(H_i H_i H_i) &=
\begin{aligned}[t]
 &3 \lambda_1  v  R^3_{i1}
				  + 3 \lambda_6  v_s  R^3_{i3}\\
			    & + 3 \lambda_7 \left( v R_{i1} R^2_{i3} + v_s   R^2_{i1}R_{i3}   \right)\,,
\end{aligned}\\
g(H_i H_j H_j) &=
\begin{aligned}[t]
 &3 \lambda_1  v   R_{i1}   R^2_{j1}
				  + 3 \lambda_6  v_s   R_{i3}   R^2_{j3} \\
			    & + \lambda_7 \left[v   \left( R_{i1}   R^2_{j3} + 2 R_{i3}   R_{j1}   R_{j3}\right)\right.\\
			    &  \left.\hphantom{+\lambda_7[} + v_s   \left(R_{i3}   R^2_{j1} + 2 R_{i1}   R_{j1}   R_{j3}\right)\right]\,,
\end{aligned}\\
g(H_i H_D H_D) &=
\frac{2}{v}\, \left(m_{H_{D}}^{2}-m_{22}^{2}\right) R_{i1} + \lambda_8 \, \frac{v_s}{v} \,\left(v    R_{i3} - v_s   R_{i1}\right)\,,\\
g(H_i H_D^{+} H_D^{-}) &= \dfrac{2}{v}\,\left(m_{H^{\pm}_D}^{2}-m_{22}^{2}\right) R_{i1} + \lambda_8\, \dfrac{v_s}{v}\,\left(v  R_{i3}-v_s  R_{i1}\right)\,,\\
g(H_i A_D A_D) &= \,\dfrac{2}{v}\,\left(m_{A_D}^{2}-m_{22}^{2}\right) R_{i1} + \lambda_8\, \dfrac{v_s}{v}\,\left(v  R_{i3}-v_s  R_{i1}\right).
\end{align}
where $H_{i,j}\in\{H_1,H_2\}$ and $R_{ij}$ is given by \cref{eq:mixmatIDP}.

\section*{Acknowledgements}
We are grateful to Marco Sampaio and Rui Santos for helpful discussions and feedback.
We thank Michael Spira for providing and maintaining the code \texttt{HDECAY} that \texttt{N2HDECAY} is based upon.
JW acknowledges funding from the ``PIER Helmholtz Graduate School''.

\section*{Bibliography}
\bibliographystyle{elsarticle-num}
\bibliography{Bibliography}

\begin{thebibliography}{10}
\expandafter\ifx\csname url\endcsname\relax
  \def\url#1{\texttt{#1}}\fi
\expandafter\ifx\csname urlprefix\endcsname\relax\def\urlprefix{URL }\fi
\expandafter\ifx\csname href\endcsname\relax
  \def\href#1#2{#2} \def\path#1{#1}\fi

\bibitem{Aad:2012tfa}
G.~Aad, et~al., {Observation of a new particle in the search for the Standard
  Model Higgs boson with the ATLAS detector at the LHC}, Phys. Lett. B716
  (2012) 1--29.
\newblock \href {http://arxiv.org/abs/1207.7214} {\path{arXiv:1207.7214}},
  \href {http://dx.doi.org/10.1016/j.physletb.2012.08.020}
  {\path{doi:10.1016/j.physletb.2012.08.020}}.

\bibitem{Chatrchyan:2012xdj}
S.~Chatrchyan, et~al., {Observation of a new boson at a mass of 125 GeV with
  the CMS experiment at the LHC}, Phys. Lett. B716 (2012) 30--61.
\newblock \href {http://arxiv.org/abs/1207.7235} {\path{arXiv:1207.7235}},
  \href {http://dx.doi.org/10.1016/j.physletb.2012.08.021}
  {\path{doi:10.1016/j.physletb.2012.08.021}}.

\bibitem{Lee:1973iz}
T.~D. Lee, {A Theory of Spontaneous T Violation}, Phys. Rev. D8 (1973)
  1226--1239, [,516(1973)].
\newblock \href {http://dx.doi.org/10.1103/PhysRevD.8.1226}
  {\path{doi:10.1103/PhysRevD.8.1226}}.

\bibitem{Branco:2011iw}
G.~C. Branco, P.~M. Ferreira, L.~Lavoura, M.~N. Rebelo, M.~Sher, J.~P. Silva,
  {Theory and phenomenology of two-Higgs-doublet models}, Phys. Rept. 516
  (2012) 1--102.
\newblock \href {http://arxiv.org/abs/1106.0034} {\path{arXiv:1106.0034}},
  \href {http://dx.doi.org/10.1016/j.physrep.2012.02.002}
  {\path{doi:10.1016/j.physrep.2012.02.002}}.

\bibitem{Chen:2013jvg}
C.-Y. Chen, M.~Freid, M.~Sher, {Next-to-minimal two Higgs doublet model}, Phys.
  Rev. D89~(7) (2014) 075009.
\newblock \href {http://arxiv.org/abs/1312.3949} {\path{arXiv:1312.3949}},
  \href {http://dx.doi.org/10.1103/PhysRevD.89.075009}
  {\path{doi:10.1103/PhysRevD.89.075009}}.

\bibitem{Drozd:2014yla}
A.~Drozd, B.~Grzadkowski, J.~F. Gunion, Y.~Jiang, {Extending two-Higgs-doublet
  models by a singlet scalar field - the Case for Dark Matter}, JHEP 11 (2014)
  105.
\newblock \href {http://arxiv.org/abs/1408.2106} {\path{arXiv:1408.2106}},
  \href {http://dx.doi.org/10.1007/JHEP11(2014)105}
  {\path{doi:10.1007/JHEP11(2014)105}}.

\bibitem{Muhlleitner:2016mzt}
M.~M{\"{u}}hlleitner, M.~O.~P. Sampaio, R.~Santos, J.~Wittbrodt, {The N2HDM
  under Theoretical and Experimental Scrutiny}, JHEP 03 (2017) 094.
\newblock \href {http://arxiv.org/abs/1612.01309} {\path{arXiv:1612.01309}},
  \href {http://dx.doi.org/10.1007/JHEP03(2017)094}
  {\path{doi:10.1007/JHEP03(2017)094}}.

\bibitem{Curtin:2013fra}
D.~Curtin, et~al., {Exotic decays of the 125{~}GeV Higgs boson}, Phys. Rev.
  D90~(7) (2014) 075004.
\newblock \href {http://arxiv.org/abs/1312.4992} {\path{arXiv:1312.4992}},
  \href {http://dx.doi.org/10.1103/PhysRevD.90.075004}
  {\path{doi:10.1103/PhysRevD.90.075004}}.

\bibitem{Athron:2017kgt}
P.~Athron, et~al., {Status of the scalar singlet dark matter model}, Eur. Phys.
  J. C77~(8) (2017) 568.
\newblock \href {http://arxiv.org/abs/1705.07931} {\path{arXiv:1705.07931}},
  \href {http://dx.doi.org/10.1140/epjc/s10052-017-5113-1}
  {\path{doi:10.1140/epjc/s10052-017-5113-1}}.

\bibitem{Deshpande:1977rw}
N.~G. Deshpande, E.~Ma, {Pattern of Symmetry Breaking with Two Higgs Doublets},
  Phys. Rev. D18 (1978) 2574.
\newblock \href {http://dx.doi.org/10.1103/PhysRevD.18.2574}
  {\path{doi:10.1103/PhysRevD.18.2574}}.

\bibitem{Ma:2006km}
E.~Ma, {Verifiable radiative seesaw mechanism of neutrino mass and dark
  matter}, Phys. Rev. D73 (2006) 077301.
\newblock \href {http://arxiv.org/abs/hep-ph/0601225}
  {\path{arXiv:hep-ph/0601225}}, \href
  {http://dx.doi.org/10.1103/PhysRevD.73.077301}
  {\path{doi:10.1103/PhysRevD.73.077301}}.

\bibitem{LopezHonorez:2006gr}
L.~Lopez~Honorez, E.~Nezri, J.~F. Oliver, M.~H.~G. Tytgat, {The Inert Doublet
  Model: An Archetype for Dark Matter}, JCAP 0702 (2007) 028.
\newblock \href {http://arxiv.org/abs/hep-ph/0612275}
  {\path{arXiv:hep-ph/0612275}}, \href
  {http://dx.doi.org/10.1088/1475-7516/2007/02/028}
  {\path{doi:10.1088/1475-7516/2007/02/028}}.

\bibitem{Belyaev:2016lok}
A.~Belyaev, G.~Cacciapaglia, I.~P. Ivanov, F.~Rojas-Abatte, M.~Thomas, {Anatomy
  of the Inert Two Higgs Doublet Model in the light of the LHC and non-LHC Dark
  Matter Searches}, Phys. Rev. D97~(3) (2018) 035011.
\newblock \href {http://arxiv.org/abs/1612.00511} {\path{arXiv:1612.00511}},
  \href {http://dx.doi.org/10.1103/PhysRevD.97.035011}
  {\path{doi:10.1103/PhysRevD.97.035011}}.

\bibitem{Ilnicka:2015jba}
A.~Ilnicka, M.~Krawczyk, T.~Robens, {Inert Doublet Model in light of LHC Run I
  and astrophysical data}, Phys. Rev. D93~(5) (2016) 055026.
\newblock \href {http://arxiv.org/abs/1508.01671} {\path{arXiv:1508.01671}},
  \href {http://dx.doi.org/10.1103/PhysRevD.93.055026}
  {\path{doi:10.1103/PhysRevD.93.055026}}.

\bibitem{Engeln:2018ywp}
I.~Engeln, {Phenomenological Comparison of the Dark Phases of the
Next-to-Two-Higgs-Doublet Model},
  Master's thesis, KIT, Karlsruhe (2018-03-07).
\newline\urlprefix\url{https://www.itp.kit.edu/_media/publications/masterthesis_isabellengeln.pdf}

\bibitem{Djouadi:1997yw}
A.~Djouadi, J.~Kalinowski, M.~Spira, {HDECAY: A Program for Higgs boson decays
  in the standard model and its supersymmetric extension}, Comput. Phys.
  Commun. 108 (1998) 56--74.
\newblock \href {http://arxiv.org/abs/hep-ph/9704448}
  {\path{arXiv:hep-ph/9704448}}, \href
  {http://dx.doi.org/10.1016/S0010-4655(97)00123-9}
  {\path{doi:10.1016/S0010-4655(97)00123-9}}.

\bibitem{Butterworth:2010ym}
J.~M. Butterworth, et~al., {THE
  TOOLS AND MONTE CARLO WORKING GROUP Summary Report from the Les Houches 2009
  Workshop on TeV Colliders}, in: {Physics at TeV colliders. Proceedings,
  6\textsuperscript{th} Workshop, dedicated to Thomas Binoth, Les Houches,
  France, June 8-26, 2009}, 2010.
\newblock \href {http://arxiv.org/abs/1003.1643} {\path{arXiv:1003.1643}}.
\newline\urlprefix\url{http://inspirehep.net/record/848006/files/arXiv:1003.1643.pdf}

\bibitem{Djouadi:2018xqq}
A.~Djouadi, J.~Kalinowski, M.~Muehlleitner, M.~Spira, {HDECAY: Twenty$_{++}$
  Years After. }\href {http://arxiv.org/abs/1801.09506}
  {\path{arXiv:1801.09506}}.

\bibitem{Harlander:2013qxa}
R.~Harlander, M.~M{\"{u}}hlleitner, J.~Rathsman, M.~Spira, O.~St{{\aa}}l,
  {Interim recommendations for the evaluation of Higgs production cross
  sections and branching ratios at the LHC in the Two-Higgs-Doublet Model.
  }\href {http://arxiv.org/abs/1312.5571} {\path{arXiv:1312.5571}}.

\bibitem{Krause:2017mal}
M.~Krause, D.~Lopez-Val, M.~Muhlleitner, R.~Santos, {Gauge-independent
  Renormalization of the N2HDM}, JHEP 12 (2017) 077.
\newblock \href {http://arxiv.org/abs/1708.01578} {\path{arXiv:1708.01578}},
  \href {http://dx.doi.org/10.1007/JHEP12(2017)077}
  {\path{doi:10.1007/JHEP12(2017)077}}.

\bibitem{Braaten:1980yq}
E.~Braaten, J.~P. Leveille, {Higgs Boson Decay and the Running Mass}, Phys.
  Rev. D22 (1980) 715.
\newblock \href {http://dx.doi.org/10.1103/PhysRevD.22.715}
  {\path{doi:10.1103/PhysRevD.22.715}}.

\bibitem{Sakai:1980fa}
N.~Sakai, {Perturbative QCD Corrections to the Hadronic Decay Width of the
  Higgs Boson}, Phys. Rev. D22 (1980) 2220.
\newblock \href {http://dx.doi.org/10.1103/PhysRevD.22.2220}
  {\path{doi:10.1103/PhysRevD.22.2220}}.

\bibitem{Inami:1980qp}
T.~Inami, T.~Kubota, {Renormalization Group Estimate of the Hadronic Decay
  Width of the Higgs Boson}, Nucl. Phys. B179 (1981) 171--188.
\newblock \href {http://dx.doi.org/10.1016/0550-3213(81)90253-4}
  {\path{doi:10.1016/0550-3213(81)90253-4}}.

\bibitem{Drees:1989du}
M.~Drees, K.-i. Hikasa, {Heavy Quark Thresholds in Higgs Physics}, Phys. Rev.
  D41 (1990) 1547.
\newblock \href {http://dx.doi.org/10.1103/PhysRevD.41.1547}
  {\path{doi:10.1103/PhysRevD.41.1547}}.

\bibitem{Drees:1990dq}
M.~Drees, K.-i. Hikasa, {NOTE ON QCD CORRECTIONS TO HADRONIC HIGGS DECAY},
  Phys. Lett. B240 (1990) 455, [Erratum: Phys. Lett.B262,497(1991)].
\newblock \href {http://dx.doi.org/10.1016/0370-2693(90)91130-4}
  {\path{doi:10.1016/0370-2693(90)91130-4}}.

\bibitem{Gorishnii:1983cu}
S.~G. Gorishnii, A.~L. Kataev, S.~A. Larin, {The Width of Higgs Boson Decay
  Into Hadrons: Three Loop Corrections of Strong Interactions}, Sov. J. Nucl.
  Phys. 40 (1984) 329--334, [Yad. Fiz.40,517(1984)].

\bibitem{Gorishnii:1990zu}
S.~G. Gorishnii, A.~L. Kataev, S.~A. Larin, L.~R. Surguladze, {Corrected Three
  Loop {QCD} Correction to the Correlator of the Quark Scalar Currents and
  $\gamma$ (Tot) ($H^0 \to$ Hadrons)}, Mod. Phys. Lett. A5 (1990) 2703--2712.
\newblock \href {http://dx.doi.org/10.1142/S0217732390003152}
  {\path{doi:10.1142/S0217732390003152}}.

\bibitem{Gorishnii:1991zr}
S.~G. Gorishnii, A.~L. Kataev, S.~A. Larin, L.~R. Surguladze, {Scheme
  dependence of the next to next-to-leading QCD corrections to Gamma(tot) (H0
  $\to$ hadrons) and the spurious QCD infrared fixed point}, Phys. Rev. D43
  (1991) 1633--1640.
\newblock \href {http://dx.doi.org/10.1103/PhysRevD.43.1633}
  {\path{doi:10.1103/PhysRevD.43.1633}}.

\bibitem{Kataev:1993be}
A.~L. Kataev, V.~T. Kim, {The Effects of the QCD corrections to Gamma (H0 $\to$
  b anti-b)}, Mod. Phys. Lett. A9 (1994) 1309--1326.
\newblock \href {http://dx.doi.org/10.1142/S0217732394001131}
  {\path{doi:10.1142/S0217732394001131}}.

\bibitem{Surguladze:1994gc}
L.~R. Surguladze, {Quark mass effects in fermionic decays of the Higgs boson in
  O (alpha-s**2) perturbative QCD}, Phys. Lett. B341 (1994) 60--72.
\newblock \href {http://arxiv.org/abs/hep-ph/9405325}
  {\path{arXiv:hep-ph/9405325}}, \href
  {http://dx.doi.org/10.1016/0370-2693(94)01253-9}
  {\path{doi:10.1016/0370-2693(94)01253-9}}.

\bibitem{Larin:1995sq}
S.~A. Larin, T.~van Ritbergen, J.~A.~M. Vermaseren, {The Large top quark mass
  expansion for Higgs boson decays into bottom quarks and into gluons}, Phys.
  Lett. B362 (1995) 134--140.
\newblock \href {http://arxiv.org/abs/hep-ph/9506465}
  {\path{arXiv:hep-ph/9506465}}, \href
  {http://dx.doi.org/10.1016/0370-2693(95)01192-S}
  {\path{doi:10.1016/0370-2693(95)01192-S}}.

\bibitem{Melnikov:1995yp}
K.~Melnikov, {Two loop O(N(f) alpha-s**2) correction to the decay width of the
  Higgs boson to two massive fermions}, Phys. Rev. D53 (1996) 5020--5027.
\newblock \href {http://arxiv.org/abs/hep-ph/9511310}
  {\path{arXiv:hep-ph/9511310}}, \href
  {http://dx.doi.org/10.1103/PhysRevD.53.5020}
  {\path{doi:10.1103/PhysRevD.53.5020}}.

\bibitem{Chetyrkin:1995pd}
K.~G. Chetyrkin, A.~Kwiatkowski, {Second order QCD corrections to scalar and
  pseudoscalar Higgs decays into massive bottom quarks}, Nucl. Phys. B461
  (1996) 3--18.
\newblock \href {http://arxiv.org/abs/hep-ph/9505358}
  {\path{arXiv:hep-ph/9505358}}, \href
  {http://dx.doi.org/10.1016/0550-3213(95)00616-8}
  {\path{doi:10.1016/0550-3213(95)00616-8}}.

\bibitem{Chetyrkin:1996sr}
K.~G. Chetyrkin, {Correlator of the quark scalar currents and Gamma(tot) (H
  $\to$ hadrons) at O (alpha-s**3) in pQCD}, Phys. Lett. B390 (1997) 309--317.
\newblock \href {http://arxiv.org/abs/hep-ph/9608318}
  {\path{arXiv:hep-ph/9608318}}, \href
  {http://dx.doi.org/10.1016/S0370-2693(96)01368-8}
  {\path{doi:10.1016/S0370-2693(96)01368-8}}.

\bibitem{Baikov:2005rw}
P.~A. Baikov, K.~G. Chetyrkin, J.~H. Kuhn, {Scalar correlator at
  O(alpha(s)**4), Higgs decay into b-quarks and bounds on the light quark
  masses}, Phys. Rev. Lett. 96 (2006) 012003.
\newblock \href {http://arxiv.org/abs/hep-ph/0511063}
  {\path{arXiv:hep-ph/0511063}}, \href
  {http://dx.doi.org/10.1103/PhysRevLett.96.012003}
  {\path{doi:10.1103/PhysRevLett.96.012003}}.

\bibitem{Inami:1982xt}
T.~Inami, T.~Kubota, Y.~Okada, {Effective Gauge Theory and the Effect of Heavy
  Quarks in Higgs Boson Decays}, Z. Phys. C18 (1983) 69.
\newblock \href {http://dx.doi.org/10.1007/BF01571710}
  {\path{doi:10.1007/BF01571710}}.

\bibitem{Djouadi:1991tka}
A.~Djouadi, M.~Spira, P.~M. Zerwas, {Production of Higgs bosons in proton
  colliders: QCD corrections}, Phys. Lett. B264 (1991) 440--446.
\newblock \href {http://dx.doi.org/10.1016/0370-2693(91)90375-Z}
  {\path{doi:10.1016/0370-2693(91)90375-Z}}.

\bibitem{Spira:1993bb}
M.~Spira, A.~Djouadi, D.~Graudenz, P.~M. Zerwas, {SUSY Higgs production at
  proton colliders}, Phys. Lett. B318 (1993) 347--353.
\newblock \href {http://dx.doi.org/10.1016/0370-2693(93)90138-8}
  {\path{doi:10.1016/0370-2693(93)90138-8}}.

\bibitem{Spira:1995rr}
M.~Spira, A.~Djouadi, D.~Graudenz, P.~M. Zerwas, {Higgs boson production at the
  LHC}, Nucl. Phys. B453 (1995) 17--82.
\newblock \href {http://arxiv.org/abs/hep-ph/9504378}
  {\path{arXiv:hep-ph/9504378}}, \href
  {http://dx.doi.org/10.1016/0550-3213(95)00379-7}
  {\path{doi:10.1016/0550-3213(95)00379-7}}.

\bibitem{Chetyrkin:1997iv}
K.~G. Chetyrkin, B.~A. Kniehl, M.~Steinhauser, {Hadronic Higgs decay to order
  alpha-s**4}, Phys. Rev. Lett. 79 (1997) 353--356.
\newblock \href {http://arxiv.org/abs/hep-ph/9705240}
  {\path{arXiv:hep-ph/9705240}}, \href
  {http://dx.doi.org/10.1103/PhysRevLett.79.353}
  {\path{doi:10.1103/PhysRevLett.79.353}}.

\bibitem{Chetyrkin:1997un}
K.~G. Chetyrkin, B.~A. Kniehl, M.~Steinhauser, {Decoupling relations to O
  (alpha-s**3) and their connection to low-energy theorems}, Nucl. Phys. B510
  (1998) 61--87.
\newblock \href {http://arxiv.org/abs/hep-ph/9708255}
  {\path{arXiv:hep-ph/9708255}}, \href
  {http://dx.doi.org/10.1016/S0550-3213(98)81004-3}{\path{doi:10.1016/S0550-3213(98)81004-3}}.

\bibitem{Kramer:1996iq}
M.~Kramer, E.~Laenen, M.~Spira, {Soft gluon radiation in Higgs boson production
  at the LHC}, Nucl. Phys. B511 (1998) 523--549.
\newblock \href {http://arxiv.org/abs/hep-ph/9611272}
  {\path{arXiv:hep-ph/9611272}}, \href
  {http://dx.doi.org/10.1016/S0550-3213(97)00679-2}
  {\path{doi:10.1016/S0550-3213(97)00679-2}}.

\bibitem{Chetyrkin:1998mw}
K.~G. Chetyrkin, B.~A. Kniehl, M.~Steinhauser, W.~A. Bardeen, {Effective QCD
  interactions of CP odd Higgs bosons at three loops}, Nucl. Phys. B535 (1998)
  3--18.
\newblock \href {http://arxiv.org/abs/hep-ph/9807241}
  {\path{arXiv:hep-ph/9807241}}, \href
  {http://dx.doi.org/10.1016/S0550-3213(98)00594-X}
  {\path{doi:10.1016/S0550-3213(98)00594-X}}.

\bibitem{Schroder:2005hy}
Y.~Schroder, M.~Steinhauser, {Four-loop decoupling relations for the strong
  coupling}, JHEP 01 (2006) 051.
\newblock \href {http://arxiv.org/abs/hep-ph/0512058}
  {\path{arXiv:hep-ph/0512058}}, \href
  {http://dx.doi.org/10.1088/1126-6708/2006/01/051}
  {\path{doi:10.1088/1126-6708/2006/01/051}}.

\bibitem{Chetyrkin:2005ia}
K.~G. Chetyrkin, J.~H. Kuhn, C.~Sturm, {QCD decoupling at four loops}, Nucl.
  Phys. B744 (2006) 121--135.
\newblock \href {http://arxiv.org/abs/hep-ph/0512060}
  {\path{arXiv:hep-ph/0512060}}, \href
  {http://dx.doi.org/10.1016/j.nuclphysb.2006.03.020}
  {\path{doi:10.1016/j.nuclphysb.2006.03.020}}.

\bibitem{Baikov:2006ch}
P.~A. Baikov, K.~G. Chetyrkin, {Top Quark Mediated Higgs Boson Decay into
  Hadrons to Order $\alpha_s^5$}, Phys. Rev. Lett. 97 (2006) 061803.
\newblock \href {http://arxiv.org/abs/hep-ph/0604194}
  {\path{arXiv:hep-ph/0604194}}, \href
  {http://dx.doi.org/10.1103/PhysRevLett.97.061803}
  {\path{doi:10.1103/PhysRevLett.97.061803}}.

\bibitem{Zheng:1990qa}
H.-Q. Zheng, D.-D. Wu, {First order QCD corrections to the decay of the Higgs
  boson into two photons}, Phys. Rev. D42 (1990) 3760--3763.
\newblock \href {http://dx.doi.org/10.1103/PhysRevD.42.3760}
  {\path{doi:10.1103/PhysRevD.42.3760}}.

\bibitem{Djouadi:1990aj}
A.~Djouadi, M.~Spira, J.~J. van~der Bij, P.~M. Zerwas, {QCD corrections to
  gamma gamma decays of Higgs particles in the intermediate mass range}, Phys.
  Lett. B257 (1991) 187--190.
\newblock \href {http://dx.doi.org/10.1016/0370-2693(91)90879-U}
  {\path{doi:10.1016/0370-2693(91)90879-U}}.

\bibitem{Djouadi:1993ji}
A.~Djouadi, M.~Spira, P.~M. Zerwas, {Two photon decay widths of Higgs
  particles}, Phys. Lett. B311 (1993) 255--260.
\newblock \href {http://arxiv.org/abs/hep-ph/9305335}
  {\path{arXiv:hep-ph/9305335}}, \href
  {http://dx.doi.org/10.1016/0370-2693(93)90564-X}
  {\path{doi:10.1016/0370-2693(93)90564-X}}.

\bibitem{Melnikov:1993tj}
K.~Melnikov, O.~I. Yakovlev, {Higgs - Two Photon Decay: QCD radiative
  correction}, Phys. Lett. B312 (1993) 179--183.
\newblock \href {http://arxiv.org/abs/hep-ph/9302281}
  {\path{arXiv:hep-ph/9302281}}, \href
  {http://dx.doi.org/10.1016/0370-2693(93)90507-E}
  {\path{doi:10.1016/0370-2693(93)90507-E}}.

\bibitem{Dawson:1992cy}
S.~Dawson, R.~P. Kauffman, {QCD corrections to H to gamma gamma}, Phys. Rev.
  D47 (1993) 1264--1267.
\newblock \href {http://dx.doi.org/10.1103/PhysRevD.47.1264}
  {\path{doi:10.1103/PhysRevD.47.1264}}.

\bibitem{Inoue:1994jq}
M.~Inoue, R.~Najima, T.~Oka, J.~Saito, {QCD corrections to two photon decay of
  the Higgs boson and its reverse process}, Mod. Phys. Lett. A9 (1994)
  1189--1194.
\newblock \href {http://dx.doi.org/10.1142/S0217732394001003}
  {\path{doi:10.1142/S0217732394001003}}.

\bibitem{Fleischer:2004vb}
J.~Fleischer, O.~V. Tarasov, V.~O. Tarasov, {Analytical result for the two loop
  QCD correction to the decay H $\to$ 2 gamma}, Phys. Lett. B584 (2004)
  294--297.
\newblock \href {http://arxiv.org/abs/hep-ph/0401090}
  {\path{arXiv:hep-ph/0401090}}, \href
  {http://dx.doi.org/10.1016/j.physletb.2004.01.063}
  {\path{doi:10.1016/j.physletb.2004.01.063}}.

\bibitem{Harlander:2005rq}
R.~Harlander, P.~Kant, {Higgs production and decay: Analytic results at
  next-to-leading order QCD}, JHEP 12 (2005) 015.
\newblock \href {http://arxiv.org/abs/hep-ph/0509189}
  {\path{arXiv:hep-ph/0509189}}, \href
  {http://dx.doi.org/10.1088/1126-6708/2005/12/015}
  {\path{doi:10.1088/1126-6708/2005/12/015}}.

\bibitem{Spira:1991tj}
M.~Spira, A.~Djouadi, P.~M. Zerwas, {QCD corrections to the H Z gamma
  coupling}, Phys. Lett. B276 (1992) 350--353.
\newblock \href {http://dx.doi.org/10.1016/0370-2693(92)90331-W}
  {\path{doi:10.1016/0370-2693(92)90331-W}}.

\bibitem{Bonciani:2015eua}
R.~Bonciani, V.~Del~Duca, H.~Frellesvig, J.~M. Henn, F.~Moriello, V.~A.
  Smirnov, {Next-to-leading order QCD corrections to the decay width H $\to
  Z\gamma$}, JHEP 08 (2015) 108.
\newblock \href {http://arxiv.org/abs/1505.00567} {\path{arXiv:1505.00567}},
  \href {http://dx.doi.org/10.1007/JHEP08(2015)108}
  {\path{doi:10.1007/JHEP08(2015)108}}.

\bibitem{Gehrmann:2015dua}
T.~Gehrmann, S.~Guns, D.~Kara, {The rare decay $H\to Z\gamma$ in perturbative
  QCD}, JHEP 09 (2015) 038.
\newblock \href {http://arxiv.org/abs/1505.00561} {\path{arXiv:1505.00561}},
  \href {http://dx.doi.org/10.1007/JHEP09(2015)038}
  {\path{doi:10.1007/JHEP09(2015)038}}.

\bibitem{Cahn:1988ru}
R.~N. Cahn, {The Higgs Boson}, Rept. Prog. Phys. 52 (1989) 389.
\newblock \href {http://dx.doi.org/10.1088/0034-4885/52/4/001}
  {\path{doi:10.1088/0034-4885/52/4/001}}.

\bibitem{Moretti:1994ds}
S.~Moretti, W.~J. Stirling, {Contributions of below threshold decays to MSSM
  Higgs branching ratios}, Phys. Lett. B347 (1995) 291--299, [Erratum: Phys.
  Lett.B366,451(1996)].
\newblock \href {http://arxiv.org/abs/hep-ph/9412209}
  {\path{arXiv:hep-ph/9412209}}, \href
  {http://dx.doi.org/10.1016/0370-2693(95)00088-3}{\path{doi:10.1016/0370-2693(95)00088-3}},
  \href{http://dx.doi.org/10.1016/0370-2693(95)01477-2}
  {\path{doi:10.1016/0370-2693(95)01477-2}}.

\bibitem{Djouadi:1995gv}
A.~Djouadi, J.~Kalinowski, P.~M. Zerwas, {Two and three-body decay modes of
  SUSY Higgs particles}, Z. Phys. C70 (1996) 435--448.
\newblock \href {http://arxiv.org/abs/hep-ph/9511342}
  {\path{arXiv:hep-ph/9511342}}, \href
  {http://dx.doi.org/10.1007/s002880050121} {\path{doi:10.1007/s002880050121}}.

\end{thebibliography}

\end{document}